# Exquisitely sensitive seal whisker-like sensors detect wakes at large distances

H.R. Beem, M.S. Triantafyllou
Department of Mechanical Engineering, Massachusetts Institute of Technology,
Cambridge, MA 02139 USA
Email address for correspondence: beem@mit.edu

**Abstract:**

Blindfolded harbor seals are able to use their uniquely shaped whiskers to track vortex wakes left by moving animals and objects that passed by up to 30 seconds earlier[1-3]; this is an impressive feat as the flow features they detect may have velocity as low as 1 mm/s, and the seals have some capacity to identify the shape of the object as well[4]. They do so while swimming forward at high speed, hence their whiskers are sensitive enough to detect small-scale changes in the external flow field, while rejecting self-generated flow noise. Here we identify and illustrate a novel flow mechanism that allows artificial whiskers with the identical unique geometry as those of the harbor seal to detect the features of minute flow fluctuations in wakes produced by objects far away. This is shown through the study of a model problem, consisting of a harbor seal whisker model interacting with the wake of an upstream circular cylinder. We show that whereas in open water the whisker geometry results in very low vibration, once it enters a wake it oscillates with large amplitude and, remarkably, its response frequency coincides with the Strouhal frequency of the upstream cylinder, thus making the detection of an upstream wake as well as an estimation of the size and shape of the wake-generating body possible. An energy flow extraction mechanism causes the large amplitude whisker oscillations to lock in to the frequency of the oncoming wake, characterized by a *slaloming* motion among the oncoming wake vortices. This passive mechanism has some similarities with the flow mechanisms observed in actively controlled propulsive foils within upstream wakes and trout swimming behind bluff cylinders in a stream, but also differences due to the remarkable whisker morphology which causes it to operate passively and within a much wider parametric range.

**Introduction:**

Behavioral experiments with harbor seals have demonstrated their outstanding ability to detect and track hydrodynamic signatures left by swimming animals and moving objects[1-3]. Despite having their auditory and visual sensory cues blocked, the seals successfully follow the paths of bodies, which had swum ahead of them by 30 seconds or longer. The seals' vibrissae (whiskers) are reported to be sensitive enough to detect the minute water movements left in these hydrodynamic trails[5].

Various animals utilize similar hair-like structures as direct touch sensors[6-8]. For aquatic animals that swim forward at relatively high speeds, however, it is rather surprising that they would be able to use their whiskers to sense hydrodynamic information. The cross-sectional shape of the whiskers is closer in form to a bluff (non-streamlined) body, whose wake is known to spontaneously form a double array of staggered vortices, the Kármán street. The unsteady forces caused by the Kármán street result in vortex-induced vibrations (VIV) on the body, with an amplitude comparable to its cross sectional dimension and with high frequency, close to the Strouhal frequency ($St = fd/U = 0.2$), resulting in high transverse velocity[9]. Such vibrations would act as strong noise for the sensor, since it would be difficult to discriminate between the wake-induced fluctuations and external flow unsteadiness.

Harbor seals and most other phocid seals have an unique whisker morphology (Fig. 1). It includes a variable cross-section in the shape of a 2:1 ellipse, whose axes vary sinusoidally in the spanwise direction, while the upstream spanwise undulation is not in phase with the downstream undulation. As found in studies of cylinders with periodic variation along the span[10-15], the whisker also experiences reduced fluid forces when the major axis of the whisker geometry is aligned with the flow[16-17]. Flow simulations reveal that these undulations break the spatial coherence of Kármán vortex formation and cause the formation of streamwise vorticity, hence reducing substantially the unsteady fluid forces on the whisker. This feature allows the seal to move forward at steady speed with small self-induced noise from VIV. The question remains on how the vibrissae detect unsteady velocities, especially the minute velocities found in wakes 30 seconds after a body has passed[3].

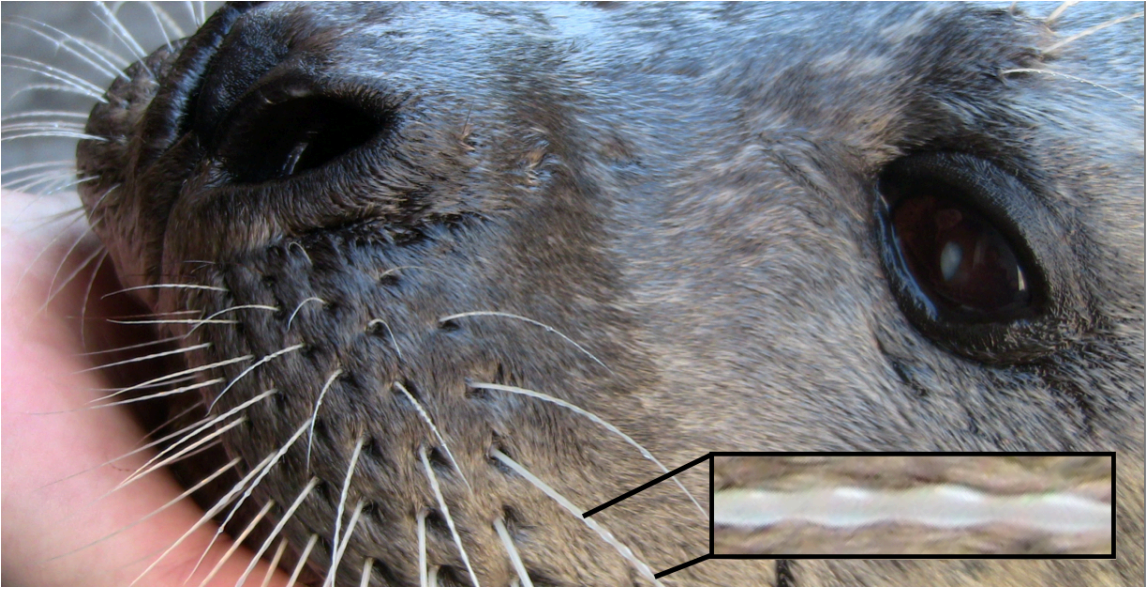

**Fig. 1. The harbor seal whisker geometry.** A close-up view of a harbor seal's muzzle reveals the whiskers' unique undulatory morphology. Photo taken courtesy of the New England Aquarium.

In order to answer this question we turn to the problem of wake-induced oscillations, which occur when a flexibly mounted body is placed in the wake of an upstream body so it is subject to the unsteady forcing of its vortical structures. Most research has addressed the problem of two interacting circular cylinders of equal or comparable diameter[18-21], in which it was shown that the upstream cylinder has significant effect on the amplitude and frequency of motion of the downstream cylinder even when it is placed 25 diameters apart. It was also found that the vortical wake of the upstream cylinder causes a restoring force on the downstream cylinder, which is equivalent to a linear spring, hence affecting its natural frequency[22]. In the few studies conducted with interfering cylinders of unequal diameter[23-25], at relatively small distances from each other, it was found that the upstream cylinder has significant effect on the response of the downstream cylinder, reducing its amplitude relative to the single cylinder case, while the frequency of oscillation of the downstream cylinder was found to be either equal to, or twice the value of the frequency of the upstream cylinder. No studies for large distances are available.

In this paper we employ a flexibly mounted model of a seal whisker placed within the wake of a larger upstream circular cylinder, modeling the response of a single seal whisker within a wake produced by an object. Strain measurements and dye visualizations are used to elucidate the whisker's mechanism of detecting the features of the unsteady flow.

**Methodology:**

*Whisker Device Design:*

A plastic whisker model (Fig. 2) was fabricated using stereolithography and mounted at its base on a flexing plate (1095 spring steel, thickness = 0.38 mm) with four arms, allowing the model to freely vibrate in the in-line (x) and crossflow (y) directions. It was therefore in a cantilevered configuration resembling that on the seal, although the main body of the whisker remained rigid in the experiment.

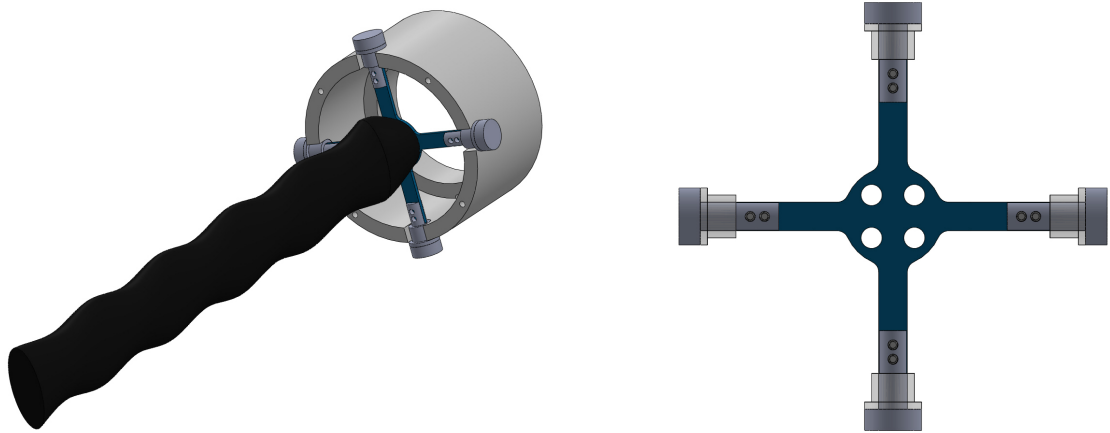

**Fig. 2. The whisker device.** A plastic whisker model is mounted on a four-armed spring steel flexure. Bushings and pins at the end of each arm allow for free rotation in the direction that is not being deflected, which reduces cross-coupling.

Strain gauges (Omega, Model KFH-03-120) on the flexing plates were arranged in a Wheatstone bridge to output a signal proportional to the deflection of the plates. Strain signals were converted to a voltage once acquired through a PhidgetBridge 1046 USB Input Board (sampling frequency = 60 Hz, gain = 16). The measurements collected were calibrated to the amount of deflection at the tip of the model. Calibration curves relating voltage and tip deflection were found to be linear for both axes on all models (an elliptical and circular cylinder of similar construction were tested for comparison), with the lowest coefficient of determination ($R^2$) value being 0.953.

The natural frequency ($f_n$) of the models was found by conducting pluck tests in water and calculating the zero-crossing frequency of the signals. The elliptical cylinder model has a natural frequency of $f_{n,x}$ = 3.1 Hz in-line and $f_{n,y}$ = 2.2 Hz in crossflow. The whisker model was designed to have similar mass, allowing it to have similar natural frequencies ($f_{n,x}$ = 3 Hz, $f_{n,y}$ = 1.8 Hz). The circular cylinder model has a smaller mass relative to the other models, giving it a lower frequency ($f_n$ = 1.8 Hz), which also necessitated the use of a slightly thinner flexure (1095 spring steel, thickness = 0.2 mm) to cover an appropriate reduced velocity range. All models have a mass ratio ($m^* = m/m_d$) of 1.4, where m is the model mass and $m_d = \rho \pi d_w^2 \text{span}/4$ is the displaced fluid mass. Due to experimental constraints, the whisker model was fabricated at a 20 times scale version of the real whisker dimensions ($d_w$ = 1.06 cm) and has a submerged span of $L/d_w$ = 30, whereas the other models are at a 30 times scale ($d_w$ = 1.59 cm) and have $L/d_w$ = 12. The Reynolds number based on the diameter, Re = $Ud_w/\nu$, ranges from 1,060 to 20,670 (given the range of U covering 0.1 to 1.3 m/s), so it is generally larger than the Reynolds number of seal whiskers which is around 1,000. The higher Reynolds number in the experiments allows easier measurement of forces and motions.

*Experimental Setup:*

The whisker model was mounted in a vertical position in the MIT Towing Tank, cantilevered at its top on a spring-like structure and facing the flow in its streamlined direction.

First, the whisker's VIV response was tested by towing it in calm water. Second, a hydrodynamic wake was generated upstream of the whisker model through the use of a vertically mounted cylinder, which shed a Kármán vortex street (Fig. 3). In the case of the seal, the whisker diameter ($d_w$), which is typically around 0.5 to 1 mm, is expected to be significantly smaller than the dimension of the dominant wake vortices, which scale with the dimension of the upstream body. The diameter of the upstream cylinder was therefore chosen to be larger than that of the whisker model, taking values $d_{cyl}$ = {2.5, 4, 11} $d_w$. The two objects were towed simultaneously along the tank at velocity U, and the whisker's amplitude and frequency of vibration were measured.

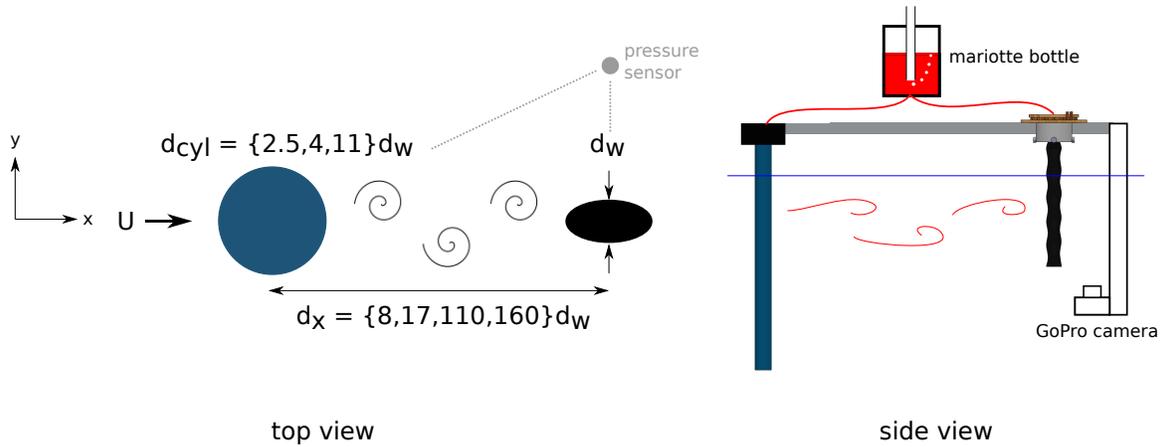

**Fig. 3. Experimental setup.** (Left) Top and (Right) side views of the setup, showing the generation of a hydrodynamic trail in front of the whisker through the use of a cylinder, which sheds a Kármán vortex street. Four different downstream distances ($d_x$) were tested, ranging from the near wake to the far wake. Three different cylinder sizes ($d_{cyl}$) were also tested. Simultaneous measurement of the oncoming wake frequency and whisker response was done through use of a pressure sensor placed near the edge of the wake. A mariotte bottle supplied dye at a constant flow rate. A camera underneath the model captured a bottom-up view of the wake-whisker interaction.

Also, the flow features of the upstream wake were measured directly and in real-time during the tests. This was done by placing a pressure sensor (Freescale MPXV7007, on-chip conditioning) at the same downstream location as the whisker to collect information on the cylinder wake. In order to avoid interfering with the flow encountered by the whisker, the pressure sensor was placed at a lateral location closer to the edge of the wake (at an angle of 60-75° between the upstream cylinder and the whisker). Plastic tubing was connected to the sensor inlet and placed into the wake, such that the sensor avoided direct contact with water. To reduce flow-induced vibrations of the tube itself, it was attached to the leading edge of a small airfoil wing, which was rigidly mounted to the carriage. A National Instruments USB-6211DAQ board was used to collect the analog signal (sampling frequency = 1000 Hz). The dominant frequency of the wake was extracted from the pressure signal and compared with that of the whisker oscillation.

A mariotte bottle setup was constructed in order to introduce dye into the flow for visualizations. This system is made of a sealed container of dye with an air inlet on the top and a dye outlet on the bottom. The air inlet ensures that the pressure inside remains constant even as the reservoir level lowers. The dye can therefore be delivered at a constant rate. The tracer was made by combining 8 parts water, 5 parts industrial food coloring, and 1 part alcohol, which resulted in a neutrally buoyant liquid.

The dye was introduced into the flow between the wake generator and the whisker. A tube from the mariotte bottle extends into the wake generator (hollow PVC pipe) and dye exits from a small hole on the downstream side of the cylinder. The hole aligns with the 65% span level of the whisker. An underwater camera captured videos of the wake-whisker interaction. A GoPro Hero 3 Black Edition was mounted below the whisker, which provided a bottom-up view. It was positioned sufficiently below and downstream of the model to avoid disturbing the flow seen by the model. Videos of the cylinder wake-whisker interaction were captured at 1440p resolution and at 48 frames per second. Snapshots of the wake behind the three geometries moving through calm water were captured at 720p resolution, 120 frames per second. In that case, the dye entered the model itself at the top and exited through a hole on the downstream side of the model again at the 65% span level.

*Analysis:*

The strain signals were recorded as function of time for each run. A bandpass Butterworth filter (pass range: 0.4 to 20 Hz) was applied. To exclude transient phenomena, only a middle segment of these signals was considered further. The vibration amplitude for that run was determined by taking the root-mean-square (RMS) amplitude of that segment. The value was then nondimensionalized by $d_w$. The crossflow frequency of vibration was determined by taking a fast Fourier transform (FFT) of the signal, selecting the peak response frequency, and nondimensionalizing it by the model's natural frequency in water: $f_y^* = f_y/f_{n,y}$.

Finally, the data utilizing simultaneous measurements of wake pressure and strain on the whisker sensor was analyzed over a more fine time scale. First, the two signals were re-sampled to the same frequency and the following signal conditioning was applied to both signals. A bandpass Butterworth filter was used to pass frequencies between 1/2 and 2 times the expected wake frequency, based on the Strouhal relation. Additional Butterworth filters were used to notch out the natural frequencies of the foil (≈1,10,20 Hz). Cross-correlation of the two signals was conducted to quantify any similarity. This correlation was first conducted by dividing the signals by their RMS value in order to normalize their magnitudes. An initial cross-correlation was carried out to determine the value of any time delay between the two signals. The signals were shifted by a delay corresponding to the peak in this initial cross-correlation result. A second cross-correlation was then used to determine the level of synchronization between the two signals. This value was normalized by the magnitude to reach a cross-correlation coefficient (c).

**Results:**

*VIV Response:*

First, the vortex-induced vibrations of the models were captured as they each moved through calm water. Dye visualizations are shown in Fig.4. The circular and elliptical cylinder shed coherent vortices in an expected pattern. The whisker generates a less coherent wake and one which forms further downstream. Note that the whisker model here uses the geometric parameters provided in previous work[16], which includes an undulation wavelength approximately half of the value seen on real seals. A model with the larger wavelength was also tested. That visualization is not shown here, but is qualitatively similar.

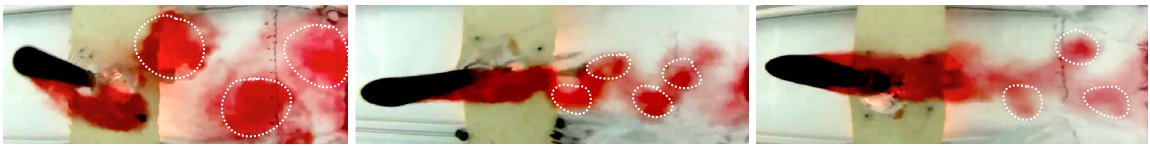

**Fig. 4. Dye visualizations of different geometries in steady flow.** The wake behind a circular cylinder, elliptical cylinder, and whisker (from left to right) positioned in the streamlined direction is visualized using dye. The first two objects shed regular 2S patterns, whereas the whisker model sheds an incoherent wake. Images were taken during runs corresponding to each model's peak amplitude: cylinder at $U^* = 10$, elliptical cylinder at $U^* = 28$, and whisker at $U^* = 28$.

*Whisker response within a wake:*

A comparison of the whisker's vibration amplitude at the model tip, nondimensionalized by the diameter, in both calm water and within the wake, is shown in Fig. 5 versus the reduced velocity ($U^* = U/f_n d_w$). When the whisker is at a relatively close distance ($d_x = 8d_w$) from the upstream cylinder ($d_{cyl} = 2.5d_w$), the amplitude of vibration peaks at $>1.5d_w$. This value has to be contrasted with the whisker's vibration amplitude as it moves through open water (depicted with "x"), which is very small - peaking at about $0.05d_w$. Hence, for this distance from ($d_x$) and size ratio with the upstream cylinder ($d_{cyl}/d_w$), the whisker can vibrate with 30 times larger amplitude when it encounters the wake than it does in open water. At a large distance away ($d_x = 160d_w$), the peak vibration remains relatively high at >3 times larger than that in open water.

Also of note is the comparison of the whisker's frequency of response between the two cases, shown in Fig. 5, where the nondimensional frequency of vibration ($f_y^*$) is plotted versus $U^*$. The frequencies plotted here were determined by selecting the peak in the Fourier transform of the corresponding signal's time trace. As the whisker moves through open water, it oscillates at small amplitude and with a relatively constant frequency. However, when the whisker encounters the oncoming vortical wake, it oscillates at a new frequency ($f_y$) that matches the vortex shedding frequency of the upstream cylinder, as measured by the pressure sensor during the same runs. Both frequencies are nondimensionalized by the whisker's natural frequency ($f_y^* = f_y/f_{n,y}$) and plotted together. Hence the whisker synchronizes with the oncoming wake dominant frequency, which is distinct from its VIV frequency in open water. The frequency of the whisker at large distance is not shown here – it is rather depicted in an upcoming figure that depicts cross-correlation.

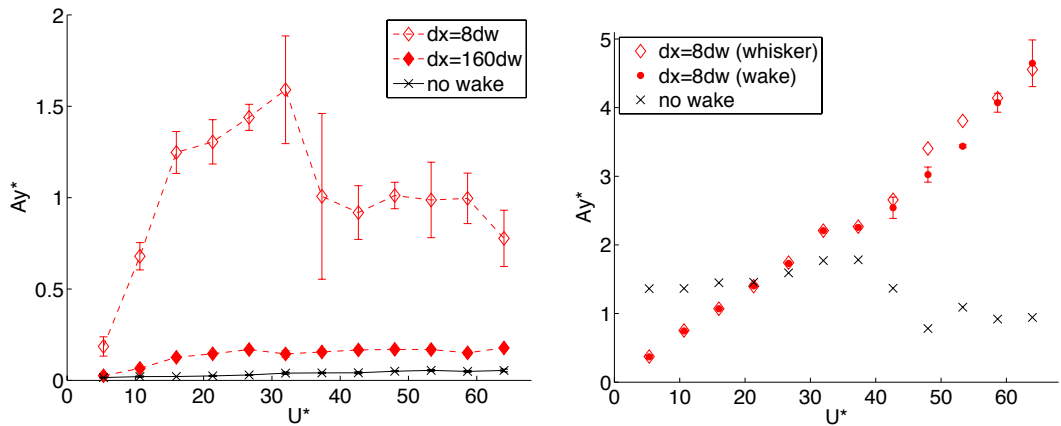

**Fig. 5. The whisker's vibration response within a wake.** For a whisker within the wake of an upstream cylinder ($d_{cyl} = 2.5d_w$) at a close distance $d_x = 8d_w$: (Left) The whisker vibrates with far higher amplitude in the wake than it does on its own. It is still significantly higher at large distances ($d_x = 160d_w$). (Right) The peak frequency of the whisker's vibration synchronizes with the oncoming wake frequency. Each symbol represents the average over three repeat runs and error bars are included.

The large amplitude amplification and Kármán frequency lock-in of the whisker constitute a mechanism of wake detection that the seals can use: The whiskers, which normally vibrate at low amplitude when in open water, vibrate with an amplitude that can be more than an order-of-magnitude larger when in the wake of a body; and, more importantly, the frequency of vibration synchronizes with the dominant frequency in the wake.

*Dye visualizations within a wake:*

Flow visualizations of the whisker interacting with the wake provide the flow mechanism that explains these findings. Videos are taken with a bottom-up perspective, under similar parameters displayed above: $d_{cyl} = 2.5d_w$, $d_x = 17d_w$, and displayed in a sequence of snapshots (Fig. 6). Dye exits the wake generator, off the right side of the images, is entrained by the vortices shed by the cylinder, and then impinges on the whisker. The whisker is painted white, while the whisker cross-section that lies within the dye plane (at 65% of the whisker span) is depicted. The principal feature is a "slaloming" trajectory, whereby the whisker comes very close to, but never pierces the nearest vortex within the Kármán street and then, as it moves further downstream, moves rapidly away from that vortex, to encounter and get close to the next vortex, at the opposite side of the Kármán street, and so on. The pressure gradient characterizing each vortex provides the suction force driving the whisker oscillations with amplitude comparable to the width of the wake, and causing synchronization with the dominant wake frequency. This constitutes the basic flow energy extraction mechanism.

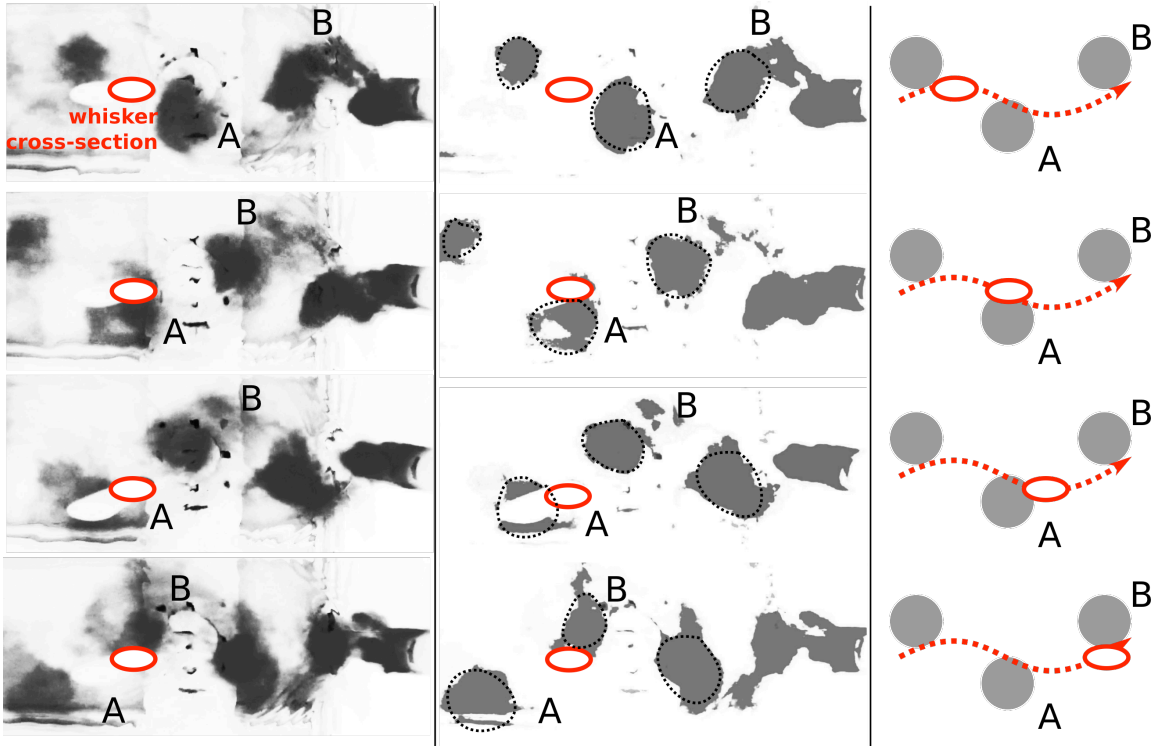

**Fig. 6. The whisker's trajectory through a vortex wake.** A sequence of dye visualizations (left: original, center: high contrast) depicting the location of the oncoming vortices relative to the cross-section of the whisker (outlined in red at the dye plane) and schematics (right) depict the whisker's free vibration path through the vortex wake. The frequency of the whisker's motion aligns with the vortex wake frequency, as the whisker slaloms between the vortices. ($U^* = 16$, Re = 3,180).

*Effects of distance and geometry:*

The amplitude values are compared in Fig. 7 for multiple distances and geometries tested. Unlike the circular cylinder, the elliptical cylinder and whisker respond with vibration amplitude detectable above the baseline VIV amplitude when a wake is present. The whisker continues to do so at large distances ($d_x > 110 d_w$).

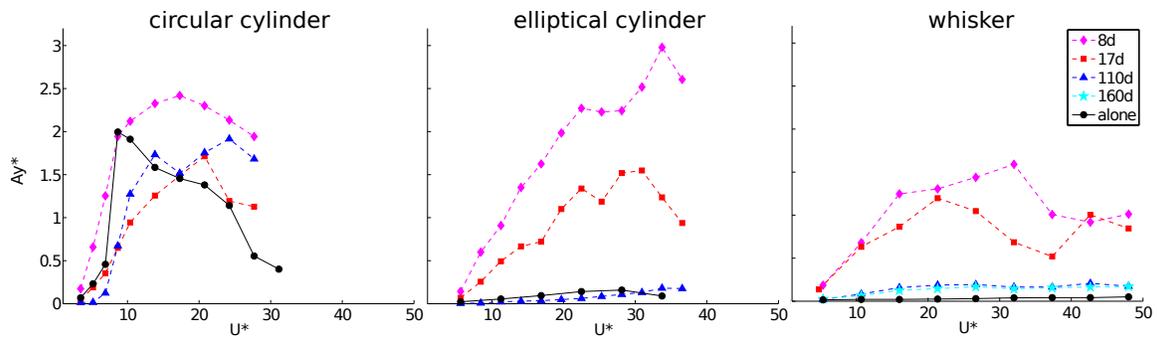

**Fig. 7. Vibration amplitudes of the circular cylinder, elliptical cylinder, and whisker.** The latter two vibrate more when they are in the wake of an upstream cylinder (colored lines) than they do in open water. The circular cylinder vibrates with similar amplitude for all cases. The amplitudes of response decrease as the distance from the upstream cylinder increases. The whisker retains an amplitude greater than its VIV amplitude even at the largest distance.

These frequencies are compared to the theoretical frequency of the wake, based on a Strouhal number of 0.2, and shown in Fig. 8 for multiple geometries and distances. At the closest distance tested, all models synchronize with the theoretical wake frequency. At a medium distance, the cylinder responds with a frequency that is consistently different than the theoretical, likely related to the wake stiffness. And at a large distance, the whisker and elliptical cylinder only align with the theoretical wake frequency over short ranges of reduced velocities.

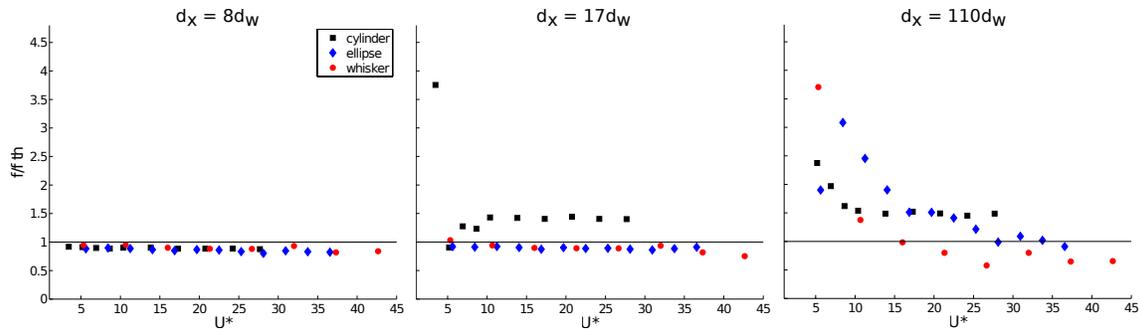

**Fig. 8. Vibration frequencies at different downstream distances.** The frequencies of response are nondimensionalized by the theoretical frequency and shown here. At the close distance ($d_x = 8dw$), all three models vibrate at the theoretical wake frequency. As the distance increases to $17d_w$, the elliptical cylinder and whisker continue to oscillate at the wake frequency, but the cylinder deviates from it. At a larger distance, only the elliptical cylinder and whisker have peak frequencies that stay close to the theoretical wake frequency over any range of reduced velocities.

*Effect of cylinder size:*

The cross-correlation coefficient is presented in Fig. 9. While Fig. 8 showed the peak frequency across the entire run to only match the expected frequency over select cases, the cross-correlation always reaches a high value. And the percent of the run that is "synchronized", defined here as having c>80% or c<-80%, and analyzed per oscillation cycle, suggests that even at large distances, the whiskers are able to lock with the frequency of the wake for some period of time that would be sufficient for wake detection. This value generally increases with the diameter of the upstream object.

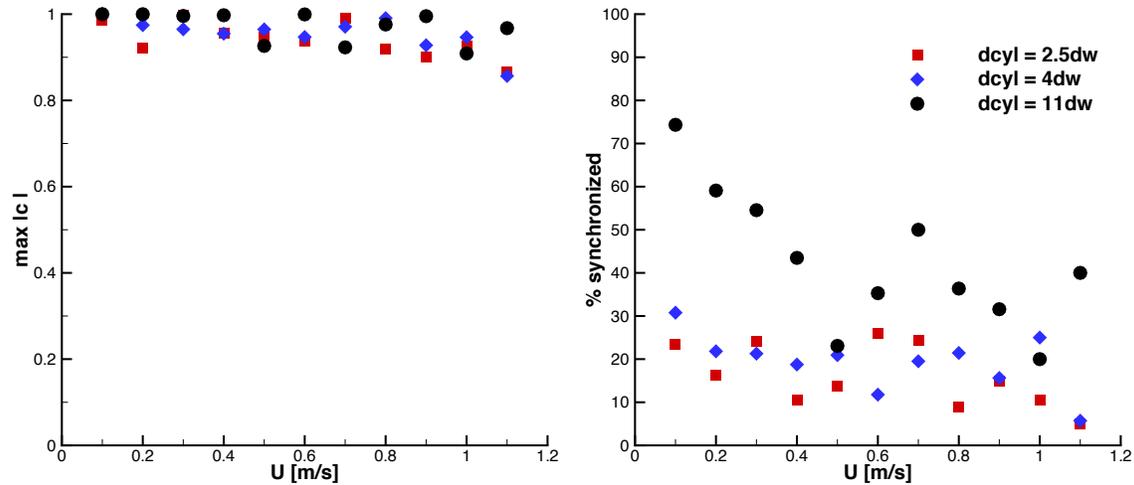

**Fig. 9. Cross-correlation metrics versus upstream cylinder diameter (at $d_x = 15d_{cyl}$).** (Left) The cross-correlation coefficient between the whisker vibration and upstream wake signals reaches a high (>0.85) maximum value for all speeds and cylinder sizes tested. (Right) The percentage of the run where the whisker is "synchronized" (defined as c>80% or c<-80%) generally increases with cylinder diameter. This indicates that the whisker's detection ability tends to increase with the size of the wake features.

**Discussion:**

A similar slaloming motion was first described for propulsive foils operating within upstream wakes, where it was shown that the slaloming trajectory leads to maximum propulsive efficiency achieved through flow energy extraction and increased thrust[26]. A basic difference between the propulsive foils and our problem is that the purpose of the former is to produce thrust at minimum energy; hence the foil generates strong vortices that combine with the oncoming Karman street vortices following the rules of vorticity control[26]. For this reason, the foil pitch angle must be controlled in order to achieve proper thrust development[26]. In contrast, the whiskers do not produce thrust and hence the vortices they shed are weak, while their pitch angle need not be controlled; therefore the flow mechanism is achieved passively and over a much wider parametric range. Another case of slaloming trajectory was found in live trout swimming behind a circular cylinder, where the fish were recorded to actively follow a roughly similar trajectory to produce thrust by extracting energy from the oncoming von Kármán vortices, also leading to synchronization with the frequency of the vortices[27]. Remarkably, an anaesthetized trout tethered in the wake of a cylinder within a cross-stream was found to synchronize passively the flapping of its body, leading to passive propulsion generating thrust through energy extraction[28]. Again, there are differences with the trout experiments, such as the formation of significant trailing edge vorticity for propulsion, and the small aspect ratio of the fish body compared to that of the oncoming vortices; however the slaloming trajectory is a basic flow energy extraction mechanism, significant to all problems.

**Conclusions:**

In the extensively studied case of two cylinders with similar diameter, the downstream cylinder oscillates at a frequency determined by its mass and spring properties, the effective added mass, and the wake spring constant. The frequency can vary substantially from both the dominant frequency of the upstream cylinder and the cylinder's natural frequency[18]. In this novel whisker-cylinder flow interaction problem, we found that the whisker vibrates with large amplitude at the Strouhal frequency of the upstream cylinder, following a path that slaloms among the vortices of the oncoming wake; if the upstream cylinder is removed, the whisker practically ceases to vibrate. Far wakes are known to reorganize their vortical structures creating new dominant frequencies[29], which places some limits on the detectability of the specific object shape, but the presence of a distinct wake is still detectable at 160 diameters downstream. We focused on a single whisker sensor, while seals have dozens of whiskers that can make detection of complex wakes feasible, but we believe that the underlying flow mechanism remains the same.

**Acknowledgments:** This work was funded in part by the Office of Naval Research under grant N00014-13-1-0059, monitored by Dr. Thomas Swean, Jr. and the Singapore National Research Foundation (NRF) through the Singapore-MIT Alliance for Research and Technology (SMART) Center for Environmental Sensing and Modeling (CENSAM). The authors would like to thank Kathy Streeter and the New England Aquarium staff for providing real whiskers shed by their harbor seals.